\documentclass[prl,twocolumn,showpacs,preprintnumbers,amsmath,amssymb,times]{revtex4}

\usepackage{graphicx}% Include figure files
\usepackage{dcolumn}% Align table columns on decimal point
\usepackage{bm}% bold math
\usepackage{psfrag}
\usepackage{epsfig}
\usepackage{amsmath}
\usepackage{amssymb}
\usepackage{color}
\usepackage{fancyheadings}
\usepackage{mathbbol}

\newcommand{\nn}{\nonumber}
\newcommand{\bra}{\langle}

\newcommand{\ket}{\rangle}

\begin{document}

\title{A polaritonic two-component Bose-Hubbard model}

\author{Michael J. Hartmann}
\email{m.hartmann@imperial.ac.uk}
\author{Fernando G.S.L. Brand\~ao}
\author{Martin B. Plenio}
\affiliation{Institute for Mathematical Sciences, Imperial College London,
53 Exhibition Road, SW7 2PG, United Kingdom}
\affiliation{QOLS, The Blackett Laboratory, Imperial College London, Prince Consort Road,
SW7 2BW, United Kingdom}

\date{\today}

\begin{abstract}
We show that polaritons in an array of interacting micro-cavities with strong atom-photon coupling
can form a two-component Bose-Hubbard model. Both polariton species are thereby protected against
spontaneous emission as their atomic part is stored in two ground states of the atoms.
The parameters of the effective model can be tuned via the driving strength of external lasers. 
We also describe a method to measure the
number statistics in one cavity for each polariton species independently. 
\end{abstract}

%Classification Scheme.
\maketitle

% ---------------------------------------------------------------------------
%
\paragraph{Introduction}

In recent years, significant progress in the theoretical and experimental study of quantum many-body
phenomena has been made by employing artificial structures that permit unprecedented experimental control and measurement access. 
Early activity in this field took place in arrays of Josephson junctions \cite{FZ01} and was followed by several important developments with ultracold atoms in optical lattices \cite{BDZ07}.
Despite their success, Josephson junction arrays and optical lattices face limitations as
it is challenging to access and control individual lattice sites, due to their small separation.

A possibility to overcome these hurdles has very recently been suggested in arrays of coupled micro-cavities,
where a scheme for simulating the Bose-Hubbard Hamiltonian \cite{HBP06} and models of
interacting Jaynes-Cummings Hamiltonians \cite{ASB06} have been proposed. The phase diagrams of
these models were studied in \cite{RF07}, where the existence of a glassy phase has been predicted.
These setups, where atoms interact with the resonant modes of the cavities, offer further possibilities to
generate effective many-body systems as they can be manipulated with external driving lasers. One of these
possibilities, effective spin Hamiltonians, has been studied recently \cite{HBP07}.

Here, we show that coupled high-Q cavities can host an effective two-component Bose-Hubbard
model,
\begin{eqnarray} \label{bosehubbard}
H_{\text{eff}} & = &
\sum_{\vec{R}; j = b,c} \mu_j \, n^{(j)}_{\vec{R}} \, - \, \sum_{\bra \vec{R}, \vec{R}' \ket; j,l = b,c} 
J_{j,l} \left( j_{\vec{R}}^{\dagger} \, l_{\vec{R}'}\, + \, \text{h.c.} \right) \, \nn \\
& + & \sum_{\vec{R}; j = b,c} U_j \, n^{(j)}_{\vec{R}} \left(n^{(j)}_{\vec{R}} - 1\right) \, 
+ \sum_{\vec{R}} U_{b,c} \, n^{(b)}_{\vec{R}} n^{(c)}_{\vec{R}} \, ,
\end{eqnarray}
where $b_{\vec{R}}^{\dagger}$($c_{\vec{R}}^{\dagger}$) create polaritons of the type $b$($c$) in the cavity at site $\vec{R}$,  $n^{(b)}_{\vec{R}} = b_{\vec{R}}^{\dagger} b_{\vec{R}}$ and
$n^{(c)}_{\vec{R}} = c_{\vec{R}}^{\dagger} c_{\vec{R}}$. $\mu_b$ and $\mu_c$ are the polariton energies, $U_b$, $U_c$ and $U_{b,c}$ their on-site interactions and $J_{b,b}$, $J_{c,c}$ and $J_{b,c}$ their tunneling rates. 

Bose-Hubbard models of two components \cite{KPS04} can display several interesting phenomena which are partly also known
for a Luttinger liquid of low energy excitations in fermionic systems. Among these are spin density separation \cite{H81,RFZZ03}, spin order in the Mott regime \cite{AHDL03} and phase separation \cite{CH03}.

We consider an array of cavities and study the dynamics of polaritons,
combined atom photon excitations, in this arrangement. Since the distance between adjacent cavities is considerably larger than the optical wavelength of the resonant mode, individual cavities can be addressed.
Photon hopping occurs between neighboring cavities while the force between two polaritons occupying the same site is generated by a large Kerr nonlinearity \cite{ISWD97}. This force can be repulsive and attractive.
Each cavity is interacting with an ensemble of these atoms, which are driven by an external laser.
By varying the intensity of the driving laser, the parameters of the effective model can be tuned.
An experimental realization would require cavities that operate in the strong coupling regime
\cite{HBW+07,BBM+05,ADW+06,SCH+06,THE+05,SKV+05}.

\paragraph{The atoms} 

\begin{figure}
\psfrag{g13}{\hspace{-0.1cm}$g_{13}$}
\psfrag{g24}{\hspace{-0.12cm}$g_{24}$}
\psfrag{o}{$\Omega$}
\psfrag{d}{\hspace{-0.06cm}$\Delta$}
\psfrag{d2}{\hspace{-0.08cm}$\delta$}
\psfrag{e}{$\varepsilon$}
\psfrag{w1}{\hspace{-0.08cm}$\omega_C$}
\psfrag{w2}{\hspace{-0.16cm}$\omega_C$}
\psfrag{1}{\raisebox{0.1cm}{$1$}}
\psfrag{2}{\raisebox{-0.02cm}{$2$}}
\psfrag{3}{$3$}
\psfrag{4}{$4$}
\includegraphics[width=6cm]{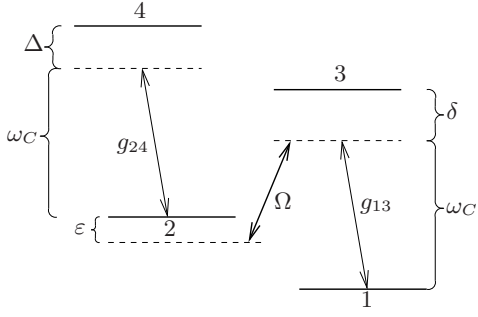}
\caption{\label{level} In each cavity atoms couple to the cavity resonance mode and an external driving laser. The figure shows the level structure and transitions of one atom, $\omega_C$ is the frequency of the cavity mode, $\Omega$ is the Rabi frequency of the driving by the laser, $g_{13}$ and $g_{24}$
are the parameters of the respective dipole couplings to the cavity mode and $\delta$, $\Delta$ and $\varepsilon$ are detunings.}
\end{figure}

To generate a force between
polaritons that are located in the same cavity, we fill the cavity with 4 level atoms,
see figure \ref{level}:
The transitions between levels 2 and 3 are coupled to the laser field
and the transitions between levels 2-4 and 1-3 couple via dipole moments to the cavity resonance mode.
It has been shown by Imamo\u{g}lu and co-workers, that this atom cavity system can show a very large nonlinearity \cite{ISWD97}.

In a rotating frame with respect to
$H_0 = \omega_C ( a^{\dagger} a + \frac{1}{2} ) + \sum_{j=1}^N ( \omega_C \sigma_{22}^j + \omega_C \sigma_{33}^j + 2 \omega_C \sigma_{44}^j )$,
the Hamiltonian of the atoms in the cavity reads,
$ H_I = \sum_{j=1}^N (\varepsilon \sigma_{22}^j + \delta \sigma_{33}^j + (\Delta + \varepsilon)
\sigma_{44}^j ) \, + \, \sum_{j=1}^N ( \Omega \, \sigma_{23}^j \, + \,
g_{13} \, \sigma_{13}^j \, a^{\dagger} \, + \,
g_{24} \, \sigma_{24}^j \, a^{\dagger} \, + \, \text{h.c.} ) \, ,$
where $\sigma_{kl}^j = | k_j \ket \bra l_j |$ transfers level $l$ of atom $j$ to level $k$ of the same atom,
$\omega_C$ is the frequency of the cavity mode, $\delta$, $\Delta$ and $\varepsilon$ are detuning parameters
(see figure \ref{level}),
$\Omega$ is the Rabi frequency of the driving by the laser and $g_{13}$ and $g_{24}$
are the parameters of the dipole coupling of the cavity mode to the respective atomic
transitions which are all assumed to be real.
All atoms interact in the same way with the cavity mode and hence the only
relevant states are Dicke type dressed states\footnote{If the atoms were distributed on fixed positions in space, the dressed states are no longer symmetric but the approach still works exactly the same.}.

\paragraph{Polaritons}

In the case where $g_{24} = 0$ and $\varepsilon = 0$ , level 4 of the atoms decouples from the dressed state
excitation manifolds \cite{ISWD97}. If we furthermore assume that the number of atoms is large, $N \gg 1$, the Hamiltonian $H_I$ can be written in terms of three polariton species. Their creation (and annihilation) operators read,
$p_0^{\dagger} = \frac{1}{B} \, \left(g S_{12}^{\dagger} - \Omega a^{\dagger} \right)$ and
$p_{\pm}^{\dagger} = \sqrt{\frac{2}{A (A \pm \delta)}} \, \left(\Omega S_{12}^{\dagger} + g a^{\dagger} \pm
\frac{A \pm \delta}{2} S_{13}^{\dagger} \right)$,
where $g = \sqrt{N} g_{13}$, $B = \sqrt{g^2 + \Omega^2}$, $A = \sqrt{4 B^2 + \delta^2}$,
$S_{12}^{\dagger} = \frac{1}{\sqrt{N}} \sum_{j=1}^N \sigma_{21}^j$
and $S_{13}^{\dagger} = \frac{1}{\sqrt{N}} \sum_{j=1}^N \sigma_{31}^j$.
The operators $p_0^{\dagger}$, $p_+^{\dagger}$ and $p_-^{\dagger}$ describe polaritons, quasi particles formed by combinations of atom and photon excitations. 

In the relevant Hilbert space spanned by symmetric Dicke type dressed states and for $N \gg 1$,
they satisfy bosonic commutation relations,
$[ p_j, p_l ] = 0$ and
$[ p_j, p_l^{\dagger} ] = \delta_{jl}$ for $j,l = 0,+,-$, 
where the neglected terms are of order $\text{''number of polaritons''}/N$.
$p_0^{\dagger}$, $p_+^{\dagger}$ and $p_-^{\dagger}$ thus describe independent bosonic
particles. In terms of these polaritons, the Hamiltonian $H_I$ for $g_{24} = \varepsilon = 0$ reads,
$\left[H_I\right]_{g_{24} = 0, \varepsilon = 0} = 
\mu_0 \, p_0^{\dagger} p_0 + \mu_+ \, p_+^{\dagger} p_+ + \mu_- \, p_-^{\dagger} p_-$,
where the frequencies are given by $\mu_0 = 0$ and $\mu_{\pm} = (\delta \pm A)/2$. 

We will now consider the case $\delta \gg \Omega, g$. Here, the polaritons and their frequencies read,
\begin{equation} \label{polariton_operators_2}
\begin{array}{rlrl}
p_0^{\dagger} = & \frac{1}{B} \, \left(g S_{12}^{\dagger} - \Omega a^{\dagger} \right) & \quad \mu_0 = & 0 \\
p_-^{\dagger} \approx & \frac{1}{B} \, \left(\Omega S_{12}^{\dagger} + g a^{\dagger} \right) - \frac{B}{\delta} S_{13}^{\dagger} & \quad \mu_- = & - \frac{B^2}{\delta} \\
p_+^{\dagger} \approx & S_{13}^{\dagger} + \frac{1}{\delta} \, \left(\Omega S_{12}^{\dagger} + g a^{\dagger} \right) & \quad \mu_+ = & \delta + \frac{B^2}{\delta}
\end{array}\, ,
\end{equation}
up to first order in $\delta^{-1}$. There is no spontaneous emission from the atomic level 2 and hence
to leading order, the polaritons $p_0^{\dagger}$ and $p_-^{\dagger}$ do not experience spontaneous emission loss. We therefore define the two polariton species
\begin{equation} \label{b_c_def}
b^{\dagger}  = \frac{1}{B} \, \left(g S_{12}^{\dagger} - \Omega a^{\dagger} \right) \, ; \quad
c^{\dagger} = \frac{1}{B} \, \left(\Omega S_{12}^{\dagger} + g a^{\dagger} \right) \, .
\end{equation}
In the rotating frame, the polaritons $b^{\dagger}$ have an energy $\mu_b = 0$ and the polaritons $c^{\dagger}$ have an energy $\mu_c = - (B/\delta) B$. A possible disorder in the resonance frequency of the cavities and hence in $\delta$ would thus affect $\mu_b$ and $\mu_c$ differently which can have interesting
consequences for the phase transitions of the model \cite{BBC+90}.
The dynamics of these two species is governed by the two component Bose-Hubbard Hamiltonian (\ref{bosehubbard}) as we shall see.

\paragraph{Perturbations} 

To write the full Hamiltonian $H_I$, in the polariton basis, we express the operators $\sum_{j=1}^N \sigma_{22}^j$ and $a^{\dagger} \, \sum_{j=1}^N \sigma_{24}^j$ in terms of $b^{\dagger}$, $c^{\dagger}$ and $p_+^{\dagger}$. We obtain,
$\sum_{j=1}^N \sigma_{42}^j \, a \approx
- S_{14}^{\dagger} \, \left(g \Omega (c^2 - b^2) + (g^2 - \Omega^2) b c \right) / B^2$,
where $S_{14}^{\dagger} = \frac{1}{\sqrt{N}} \sum_{j=1}^N \sigma_{41}^j$,
and we made use of the rotating wave approximation: Since $\delta \gg \Omega, g$, couplings to the polaritons $p_+^{\dagger}$ are negligible, provided that 
\begin{equation} \label{rotatingwappr}
|g_{24}| \, , \, |\varepsilon| \, , \, |\Delta| \, \ll \, |\mu_+ - \mu_b| \, , \, |\mu_+ - \mu_c| \, .
\end{equation}

For $\text{max}(|g_{24 }g \Omega / B^2|, |g_{24 } (g^2 - \Omega^2)/B^2|) \ll |\Delta|$, the couplings to level 4 can be treated in a perturbative way.
If furthermore $|g_{24 }g \Omega / B^2| \ll |B^2 / \delta |$, this results in energy shifts of $n_b \, (n_b - 1) \, U_b$, $n_c \, (n_c - 1) \, U_c$ and $n_b \, n_c \, U_{bc}$, where $n_b$ and $n_c$ are the numbers of $b^{\dagger}$ respectively $c^{\dagger}$ polaritons. 
The on-site interactions for the polaritons $b^{\dagger}$ and $c^{\dagger}$ can thus be written
as \footnote{For $|g_{24 }g \Omega / B^2| > |B^2 / \delta |$, an additional term $- g_{24}^2 g^2 \Omega^2 B^{-4} \Delta^{-1} (c^{\dagger} c^{\dagger} b b + b^{\dagger} b^{\dagger} c c)$ arises.}
\begin{equation} \label{Heffect_os}
U_b \, b^{\dagger}b \, ( b^{\dagger}b - 1 ) \, + \, U_c \, c^{\dagger}c \, ( c^{\dagger}c - 1 ) \, + \,
U_{bc} \, b^{\dagger}b \, c^{\dagger}c
\end{equation}
with $U_b = - g_{24}^2 g^2 \Omega^2 B^{-4} \Delta^{-1}$, $U_c = - g_{24}^2 g^2 \Omega^2 B^{-4} (\Delta + 2 B^2/\delta)^{-1}$
and $U_{bc} = - g_{24}^2 (g^2 - \Omega^2)^2 B^{-4} (\Delta + B^2/\delta)^{-1}$.
Note that $U_b > 0$ if $\Delta < 0$, $U_c > 0$ if $\Delta + 2 B^2/\delta < 0$,
$U_{bc} > 0$ if $\Delta + B^2/\delta < 0$ and vice versa. There can thus be repulsive and attractive interactions at the same time, e.g. for $\Delta < 0$ and $|\Delta| < B^2/\delta$ we have $U_b > 0$,
$U_c < 0$ and $U_{bc} < 0$.
In a similar way, the two photon detuning $\varepsilon$ leads to and additional on-site term
\begin{equation} \label{Heffect_os_2}
\frac{\varepsilon}{B^2} \left( g^2 b^{\dagger}b + \Omega^2 c^{\dagger}c + g \Omega ( b^{\dagger}c + c^{\dagger}b ) \right) \, ,
\end{equation}
where the transitions $b^{\dagger}c + c^{\dagger}b$
are suppressed if $|\varepsilon g \Omega / B^2| \ll |B^2 / \delta|$.

\paragraph{Polariton tunneling}

If the cavities are either coupled by optical fiber tapers or directly via an overlap of evanescent fields,
photons can tunnel between neighboring cavities. This process is described by the Hamiltonian
$\alpha \, (a_{\vec{R}}^{\dagger} a_{\vec{R}'} + \text{h.c.})$, where $\alpha$ is the tunneling rate of the photons. We translate this term into the polariton picture and
assume that the tunneling rate is much smaller than $\delta$. In this regime, the tunneling does not
induce transitions between the polaritons $b^{\dagger}$ or $c^{\dagger}$ and $p_+^{\dagger}$.
Hence the $p_+^{\dagger}$ decouple from the polaritons $b^{\dagger}$ and $c^{\dagger}$ whose tunneling terms read,
\begin{equation} 
J_{bb} b_{\vec{R}}^{\dagger} b_{\vec{R}'} +
J_{cc} c_{\vec{R}}^{\dagger} c_{\vec{R}'} -
J_{bc} (b_{\vec{R}}^{\dagger} c_{\vec{R}'} + c_{\vec{R}}^{\dagger} b_{\vec{R}'}) + \text{h.c.} \, ,
\end{equation}
where $J_{bb} = \alpha g^2/ B^2$, $J_{cc} = \alpha \Omega^2 / B^2$ and $J_{bc} = \alpha g \Omega / B^2$. If $|J_{bc}| \ll |B^2 / \delta|$, transitions between $b^{\dagger}$ and $c^{\dagger}$ are suppressed. This is the case for any $\Omega$ as long as $g^2 \gg \alpha \delta / 2$.

\paragraph{Parameter range}

\begin{figure}
\psfrag{xa}{\hspace{0.1cm} $\frac{\Omega}{g_{13}}$}
\psfrag{xb}{\hspace{0.1cm} $\frac{\Omega}{g_{13}}$}
\psfrag{a1}{\raisebox{-0.2cm}{\scriptsize $10^{1}$}}
\psfrag{a2}{\raisebox{-0.2cm}{\scriptsize $10^{2}$}}
\psfrag{a3}{\raisebox{-0.2cm}{\scriptsize $10^{3}$}}
\psfrag{aa1}{\hspace{-0.5cm} \raisebox{-0.0cm}{\scriptsize $10^{-3}$}}
\psfrag{aa2}{\hspace{-0.5cm} \raisebox{-0.0cm}{\scriptsize $10^{-2}$}}
\psfrag{aa3}{\hspace{-0.5cm} \raisebox{-0.0cm}{\scriptsize $10^{-1}$}}
\psfrag{aa4}{\hspace{-0.3cm} \raisebox{-0.0cm}{\scriptsize $10^{0}$}}
\psfrag{aa5}{\hspace{-0.3cm} \raisebox{-0.0cm}{\scriptsize $10^{1}$}}
\psfrag{b1}{\raisebox{-0.2cm}{\scriptsize $10^{1}$}}
\psfrag{b2}{\raisebox{-0.2cm}{\scriptsize $10^{2}$}}
\psfrag{b3}{\raisebox{-0.2cm}{\scriptsize $10^{3}$}}
\psfrag{bb1}{\hspace{-0.5cm} \raisebox{-0.0cm}{\scriptsize $10^{-4}$}}
\psfrag{bb2}{\hspace{-0.5cm} \raisebox{-0.0cm}{\scriptsize $10^{-3}$}}
\psfrag{bb3}{\hspace{-0.5cm} \raisebox{-0.0cm}{\scriptsize $10^{-2}$}}
\psfrag{bb4}{\hspace{-0.5cm} \raisebox{-0.0cm}{\scriptsize $10^{-1}$}}
\psfrag{bb5}{\hspace{-0.3cm} \raisebox{-0.0cm}{\scriptsize $10^{0}$}}
\psfrag{bb6}{\hspace{-0.3cm} \raisebox{-0.0cm}{\scriptsize $10^{1}$}}
\includegraphics[width=3.4cm]{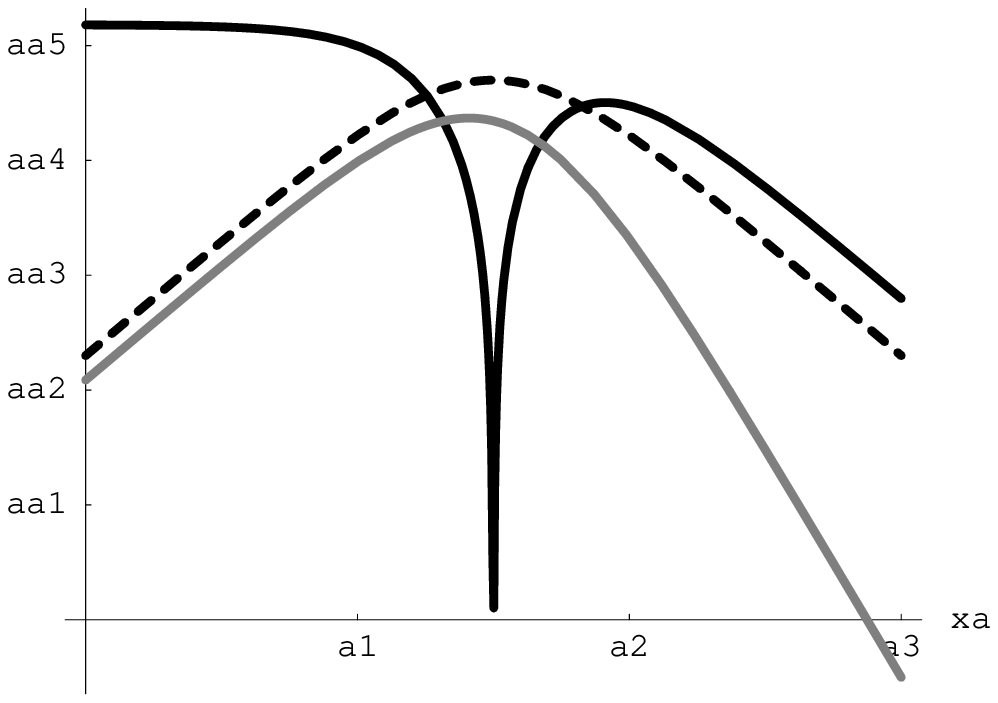}
\hspace{0.8cm}
\includegraphics[width=3.4cm]{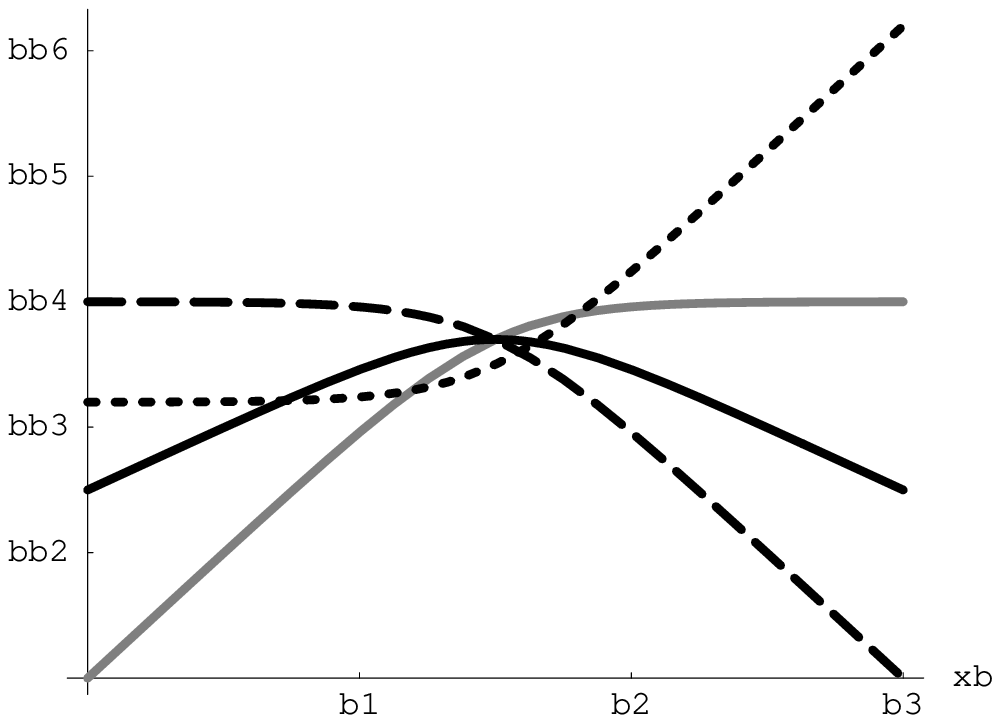}
\caption{\label{range} Left: The polariton interactions $U_b$ (dashed line), $U_c$ (gray line) and $U_{bc}$
(solid line) as a function of $\Omega / g_{13}$. Right: The tunneling rates $|J_{bb}|$ (dashed line), $|J_{cc}|$ (gray line) and $|J_{bc}|$ (solid line) together with $|\mu_c - \mu_b|$ (dotted line) as a function of $\Omega / g_{13}$. The parameters of the system are $g_{24} = g_{13}$, $N = 1000$, $\Delta = - g_{13} / 20$, $\delta = 2000 \sqrt{N} g_{13}$ and $\alpha = g_{13} / 10$.}
\end{figure}

Here we give one example how the parameters of the effective Hamiltonian (\ref{bosehubbard}) vary as a function of the intensity of the driving laser $\Omega$. We choose the parameters of the atom cavity system
to be $g_{24} = g_{13}$, $N = 1000$, $\Delta = - g_{13} / 20$, $\delta = 2000 \sqrt{N} g_{13}$ and $\alpha = g_{13} / 10$. Figure \ref{range} shows the interactions $U_b$, $U_c$ and $U_{bc}$, the tunneling rates $J_{bb}$, $J_{cc}$ and $J_{bc}$ and $|\mu_c - \mu_b|$ as a function of $\Omega / g_{13}$.
For $g \approx \Omega$ we have $|U_{bc}| \ll |U_b|, |U_c|$ and $J_{bb} \approx J_{cc} \approx J_{bc}$.
Whenever $|\mu_c - \mu_b| < |J_{bc}|$, $b^{\dagger}$ polaritons get converted into $c^{\dagger}$ polaritons
and vice versa via the tunneling $J_{bc}$.
With the present choice of $\alpha$ and $\delta$, this happens for
$0.16 g \lesssim \Omega \lesssim  1.6 g$. To avoid such processes, one either needs to choose
$\alpha$ smaller or $\delta$ larger, where both choices would require higher Q of the cavities to ensure sufficient lifetime. The interactions $U_b$, $U_c$ and $U_{bc}$ can furthermore be adjusted by varying the detuning $\Delta$. This can be done by generating a Stark shift to the atomic level 4 with an additional laser that drives the transition between level 4 and a further atomic level in a dispersive (detuned) way. 

\paragraph{Numerical results}

To confirm the validity of the approximations involved in the above derivation, we present a numerical
simulation of the full dynamics of polaritons $b^{\dagger}$ and $c^{\dagger}$ in three cavities that each couple to $N=1000$ atoms and compare it to the dynamics of the corresponding effective model (\ref{bosehubbard}). We consider initial conditions with exactly one polariton $b^{\dagger}$ in cavity 1 and in cavity 2 and exactly one polariton $c^{\dagger}$ in cavity 3.
\begin{figure}
\psfrag{t}{\raisebox{-0.4cm}{\scriptsize \hspace{-0.8cm} $t \: \text{in} \: 100 \times g_{13}^{-1}$}}
\psfrag{A}{\bf \hspace{-1.6cm} a}
\psfrag{B}{\bf \hspace{-1.6cm} b}
\psfrag{dat1}{\scriptsize $N_b$}
\psfrag{dat2}{\scriptsize $N_c$}
\psfrag{dat3}{\scriptsize $\hspace{-0.02cm} \delta \hspace{-0.02cm} N_b$}
\psfrag{dat4}{\scriptsize $\hspace{-0.02cm} \delta \hspace{-0.02cm} N_c$}
\psfrag{data1}{\scriptsize $N_b$}
\psfrag{data2}{\scriptsize $N_c$}
\psfrag{data3}{\scriptsize $\hspace{-0.02cm} \delta \hspace{-0.02cm} N_b$}
\psfrag{data4}{\scriptsize $\hspace{-0.02cm} \delta \hspace{-0.02cm} N_c$}
\psfrag{0a}{\raisebox{-0.12cm}{\scriptsize $0$}}
\psfrag{50a}{\raisebox{-0.12cm}{\scriptsize $ $}}
\psfrag{100a}{\raisebox{-0.12cm}{\scriptsize $1$}}
\psfrag{150a}{\raisebox{-0.12cm}{\scriptsize $ $}}
\psfrag{200a}{\raisebox{-0.12cm}{\scriptsize $2$}}
\psfrag{250a}{\raisebox{-0.12cm}{\scriptsize $ $}}
\psfrag{300a}{\raisebox{-0.12cm}{\scriptsize $3$}}
\psfrag{350a}{\raisebox{-0.12cm}{\scriptsize $ $}}
\psfrag{400a}{\raisebox{-0.12cm}{\scriptsize $4$}}
\psfrag{450a}{\raisebox{-0.12cm}{\scriptsize $ $}}
\psfrag{500a}{\raisebox{-0.12cm}{\scriptsize $5$}}
\psfrag{550a}{\raisebox{-0.12cm}{\scriptsize $ $}}
\psfrag{600a}{\raisebox{-0.12cm}{\scriptsize $6$}}
\psfrag{800a}{\raisebox{-0.12cm}{\scriptsize $8$}}
\psfrag{1000a}{\raisebox{-0.12cm}{\scriptsize $10$}}
\psfrag{0}{\hspace{-0.3cm} \scriptsize $0$}
\psfrag{0.1}{\hspace{-0.4cm} \scriptsize $ $}
\psfrag{0.2}{\hspace{-0.4cm} \scriptsize $0.2$}
\psfrag{0.3}{\hspace{-0.4cm} \scriptsize $ $}
\psfrag{0.4}{\hspace{-0.4cm} \scriptsize $0.4$}
\psfrag{0.5}{\hspace{-0.4cm} \scriptsize $ $}
\psfrag{0.6}{\hspace{-0.4cm} \scriptsize $0.6$}
\psfrag{0.7}{\hspace{-0.4cm} \scriptsize $ $}
\psfrag{0.8}{\hspace{-0.4cm} \scriptsize $0.8$}
\psfrag{0.9}{\hspace{-0.4cm} \scriptsize $ $}
\psfrag{1}{\hspace{-0.3cm} \scriptsize $1$}
\psfrag{-0.04}{\hspace{-0.61cm} \scriptsize $-0.04$}
\psfrag{-0.03}{\hspace{-0.61cm} \scriptsize $-0.03$}
\psfrag{-0.02}{\hspace{-0.61cm} \scriptsize $-0.02$}
\psfrag{-0.015}{\hspace{-0.4cm} \scriptsize $ $}
\psfrag{-0.01}{\hspace{-0.61cm} \scriptsize $-0.01$}
\psfrag{-0.005}{\hspace{-0.4cm} \scriptsize $ $}
\psfrag{0.005}{\hspace{-0.4cm} \scriptsize $ $}
\psfrag{0.01}{\hspace{-0.49cm} \scriptsize $0.01$}
\psfrag{0.015}{\hspace{-0.4cm} \scriptsize $ $}
\psfrag{0.02}{\hspace{-0.49cm} \scriptsize $0.02$}
\psfrag{0.03}{\hspace{-0.49cm} \scriptsize $0.03$}
\psfrag{0.04}{\hspace{-0.49cm} \scriptsize $0.04$}
\includegraphics[width=4.2cm]{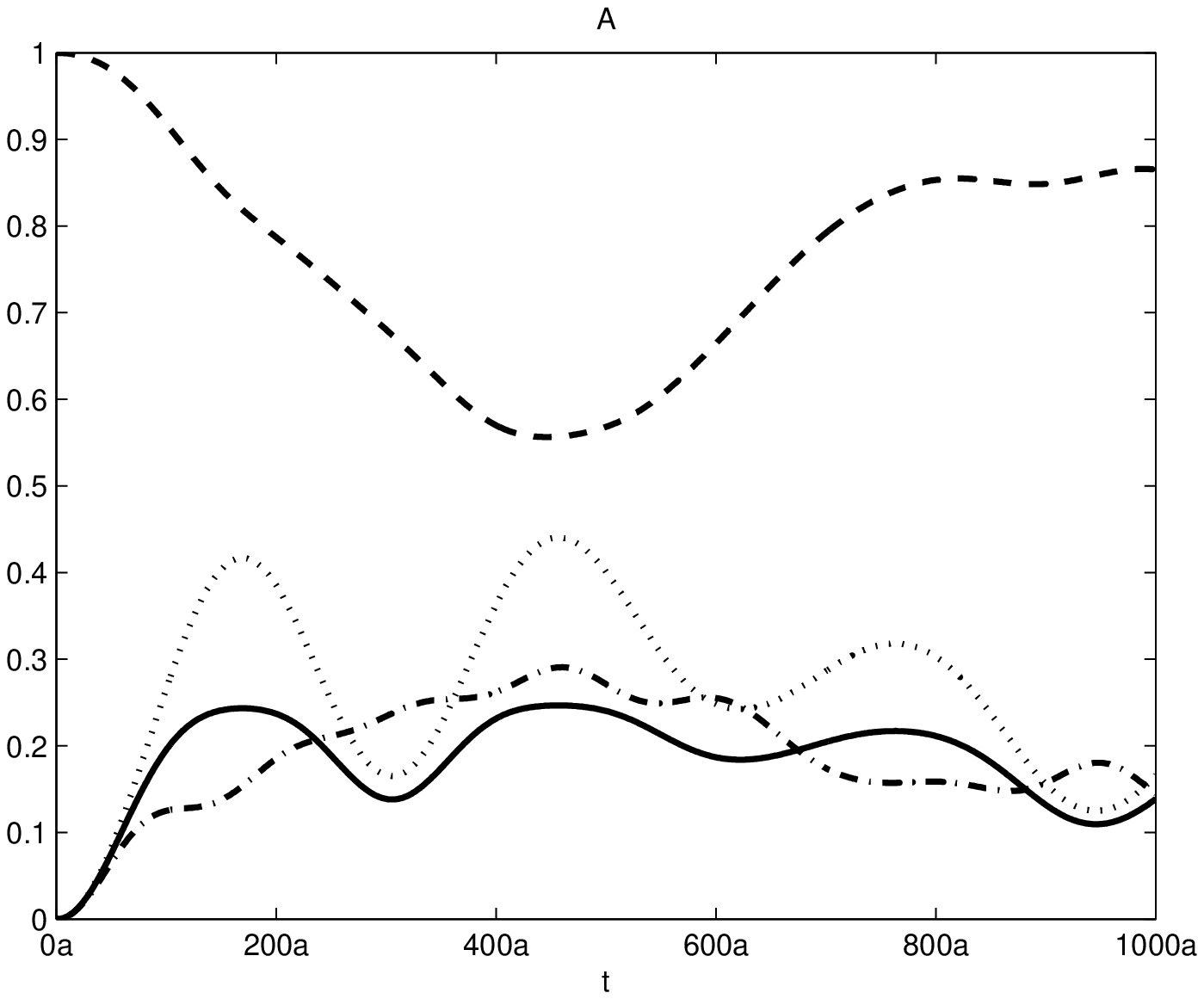}
%\includegraphics[width=4.2cm]{../numerics/3cav3/full3.eps}
%\hspace{0.2cm}
\includegraphics[width=4.2cm]{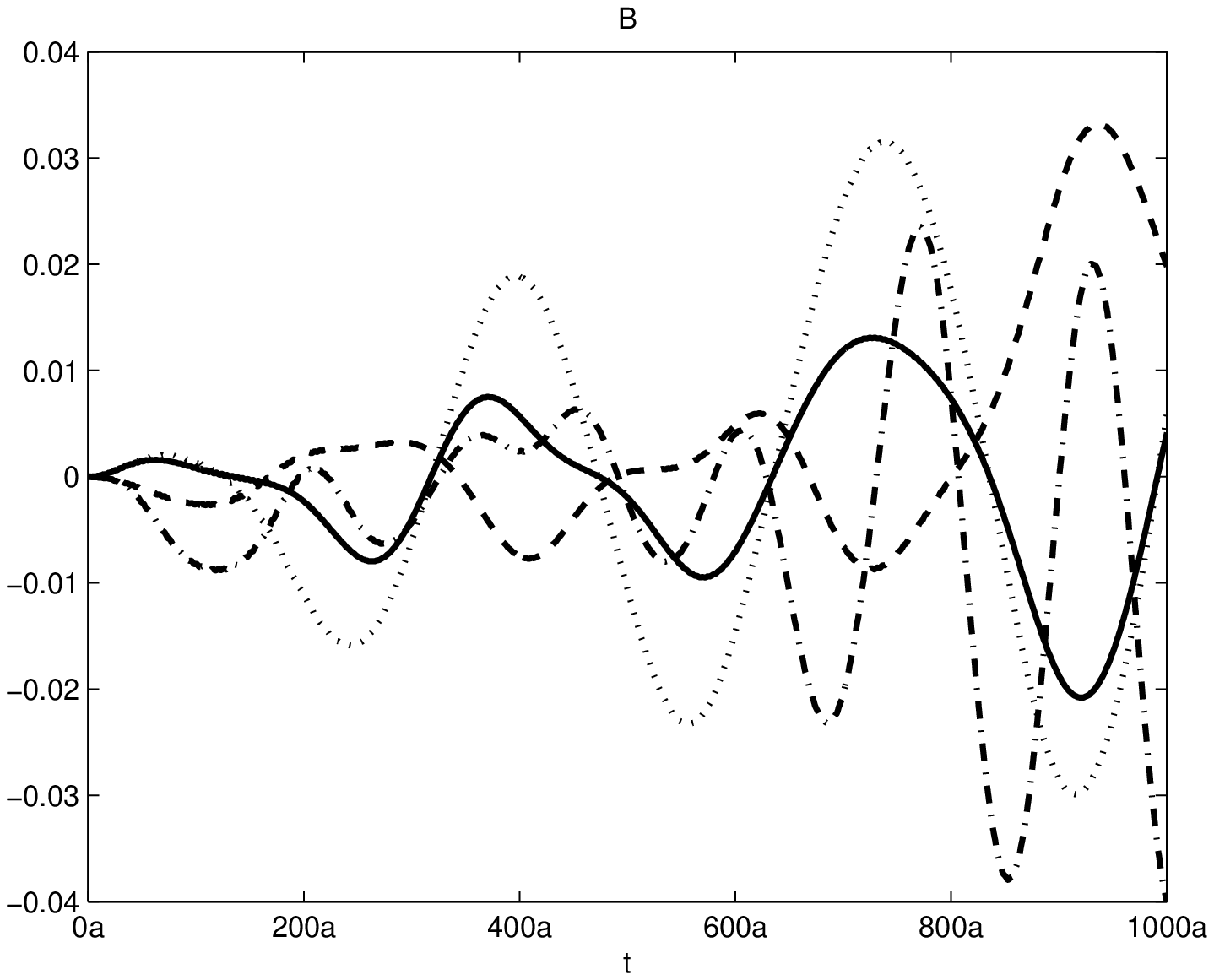}
\caption{\label{dyn} {\bf a}: $N_b$ (dotted line), $N_c$ (dashed line), $F_b$ (solid line) and $F_c$ (dash-dotted line) for a full model of 3 cavities with $g_{24} = g_{13}$,
$\varepsilon = 0$, $N = 1000$, $\Omega = \frac{3}{2} \sqrt{N} g_{13}$, $\delta = 10^4 g_{13}$,
$\Delta = - 46 g_{13}$ and $\alpha = - 2.2 \times 10^{-3} g_{13}$. {\bf b}: Differences between the full and the effective description, $[N_b]_{\text{full}} - [N_b]_{\text{BH}}$ (dotted line),
$[N_c]_{\text{full}} - [N_c]_{\text{BH}}$ (dashed line),
$[F_b]_{\text{full}} - [F_b]_{\text{BH}}$ (solid line) and
$[F_c]_{\text{full}} - [F_c]_{\text{BH}}$ (dash-dotted line) for the same model.}
\end{figure}
Figure \ref{dyn}{\bf a} shows the numbers $N_b = \bra n_b \ket$ and $N_c = \bra n_c \ket$ of polaritons $b^{\dagger}$ and $c^{\dagger}$ and their number fluctuations $F_b = \bra n_b^2 \ket - N_b^2$ and
$F_c = \bra n_c^2 \ket - N_c^2$ for the first cavity. Figure \ref{dyn}{\bf b} in turn shows differences between the full description and the effective model (\ref{bosehubbard}),
$[N_b]_{\text{full}} - [N_b]_{\text{BH}}$, $[N_c]_{\text{full}} - [N_c]_{\text{BH}}$,
$[F_b]_{\text{full}} - [F_b]_{\text{BH}}$ and
$[F_c]_{\text{full}} - [F_c]_{\text{BH}}$.
The effective model describes the dynamics very well.

\paragraph{Spontaneous emission and cavity decay}

Level 2 of the atoms is metastable and hence its decay rate negligible on the relevant time scales.
The decay mechanisms for the polaritons $b^{\dagger}$ and $c^{\dagger}$ thus originate in the cavity decay of the photons and the very small but non-negligible occupations of the excited levels 3 and 4.
The occupation of level 4 is due to the coupling $\sum_{j=1}^N (\sigma_{42}^j a + \text{h.c.})$,
whereas the occupation of level 3
only affects the polaritons $c^{\dagger}$ and stems from the linear correction term $- (B / \delta) S_{13}^{\dagger}$ in equation (\ref{polariton_operators_2}).
The resulting effective decay rates, $\Gamma_b$ for the polaritons $b^{\dagger}$ and $\Gamma_c$ for the polaritons $c^{\dagger}$, read
$\Gamma_b = \frac{\Omega^2}{B^2} \kappa  + \Theta(n_b - 2) \frac{g_{24}^2 g^2 \Omega^2}{\Delta^2 B^4} \gamma_4$
and
$\Gamma_c = \frac{g^2}{B^2} \kappa + \frac{B^2}{\delta^2} \gamma_3 + \Theta(n_c - 2) \frac{g_{24}^2 g^2 \Omega^2}{\Delta^2 B^4} \gamma_4$,
where $\Theta$ is the Heaveside step function, $\kappa$ the cavity decay rate and $\gamma_3$ ($\gamma_4$)
the spontaneous emission rates from levels 3 (4).
For successfully observing the dynamics and phases of the effective Hamiltonian
(\ref{bosehubbard}), the interactions $U_b$, $U_c$ and $U_{bc}$
need to be much larger than $\Gamma_b$ and $\Gamma_c$.

The experimentally least demanding case is the one-component model for the polaritons $b^{\dagger}$, for which $\delta \sim g$. Assuming $g_{24}= g_{13}$ the maximal achievable ratio of $U_b / \Gamma_b$ is here $\frac{1}{2} \, g_{13} / \sqrt{\kappa \, \Theta(n_b - 2) \gamma_4}$. In particular the Mott state for the polaritons $b^{\dagger}$, where $n_b \not= 2$, can even be realized in bad cavities without the strong coupling regime. However, to observe the transition to the superfluid phase, the strong coupling regime with $g_{13} \gg \sqrt{\kappa \gamma_4}$ is required for the single component model, too.

To obtain an estimate for a model with both components, $b^{\dagger}$ and $c^{\dagger}$,
we consider three cases, $g \approx \Omega$, $\Omega \approx 10 g$ and $\Omega \approx g / 10$.
Note that $g \ll \delta$ and hence spontaneous emission via level 3 is strongly suppressed.
Denoting $\zeta = g_{13} / \sqrt{\kappa \gamma_4}$, the achievable ratios of interaction versus decay rates for $g \approx \Omega$ are
$U_b / \Gamma_b \approx U_c / \Gamma_c \approx \zeta / (2 \sqrt{2})$, while the cross interaction vanishes, $U_{bc} \approx 0$.
For $\Omega = 10 g$ ($\Omega = g / 10$) the achievable ratios are
$U_b / \Gamma_b \approx \zeta / 100$ ($U_b / \Gamma_b \approx \zeta / 2$),
$U_c / \Gamma_c \approx \zeta / 2$ ($U_c / \Gamma_c \approx \zeta / 100$),
and
$U_{bc} / \text{max}(\Gamma_b,\Gamma_c) \approx \zeta$
($U_{bc} / \text{max}(\Gamma_b,\Gamma_c) \approx \zeta$).

Realizing these parameters requires cavities that operate in the strong coupling regime with large cooperativity factors, $\zeta \gg 1$. This regime is currently being achieved in several devices,
in photonic band gap cavities \cite{HBW+07} ($\zeta \approx 3$),
Fabry-Perot cavities \cite{BBM+05} ($\zeta \approx 13$),
toroidal micro-cavities \cite{ADW+06} ($\zeta \approx 7$),
fiber cavities \cite{SCH+06} ($\zeta \approx 17$) and micro-cavities on a gold coated silicon chip \cite{THE+05} ($\zeta \approx 6$) among others. Our scheme should thus be experimentally feasible with current or soon to be available technology. Values of $\zeta$ that are predicted to be achievable are as high as 200 for photonic band gap cavities and 3000 for toroidal micro-cavities \cite{SKV+05}.
Besides the strong coupling, a realization of our scheme also requires trapping the atoms in the
location of strong coupling for sufficient time.

\paragraph{Measurements}

The number statistics of both polariton species $b^{\dagger}$ and $c^{\dagger}$ in one cavity can be
measured using state selective resonance fluorescence in a way proposed in \cite{I02}. In the one-component BH model \cite{HBP06}, the polaritons can therefore be mapped by a STIRAP passage \cite{FIM05} onto the atomic levels. In the two-component case the STIRAP can however not be applied as in \cite{HBP06} because the energies $\mu_b$ and $\mu_c$ are similar and the passage would thus need to be extremely slow to still be adiabatic.

\begin{figure}
\psfrag{A}{\bf a}
\psfrag{B}{\bf b}
\psfrag{g24}{$g_{24}$}
\psfrag{o}{$\Theta$}
\psfrag{L1}{$\Lambda$}
\psfrag{L2}{$\Lambda$}
\psfrag{d}{\hspace{-0.1cm}$\Delta$}
\psfrag{d2}{\hspace{-0.0cm}$\delta_{\Lambda}$}
\psfrag{e}{$\varepsilon$}
\psfrag{w}{\hspace{-0.2cm}$\omega_C$}
\psfrag{1}{\raisebox{0.0cm}{$1$}}
\psfrag{2}{\raisebox{-0.1cm}{$2$}}
\psfrag{3}{$3$}
\psfrag{4}{$4$}
\includegraphics[width=3.4cm]{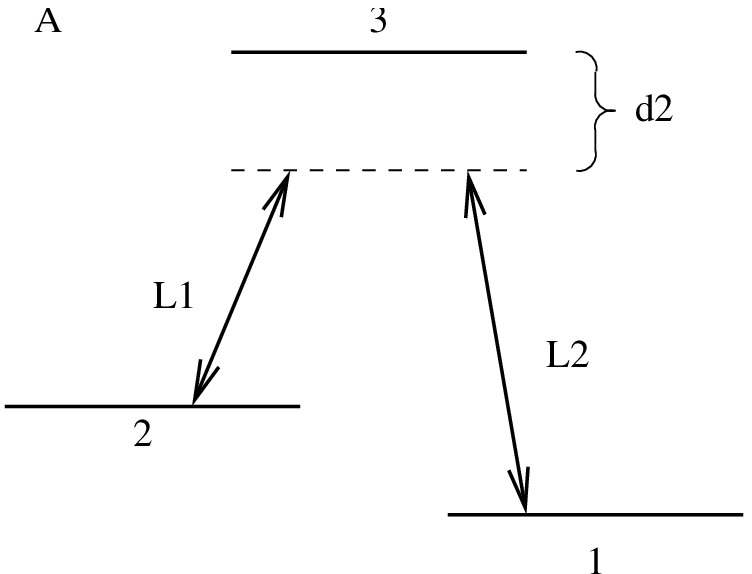}
\hspace{0.4cm}
\includegraphics[width=4.2cm]{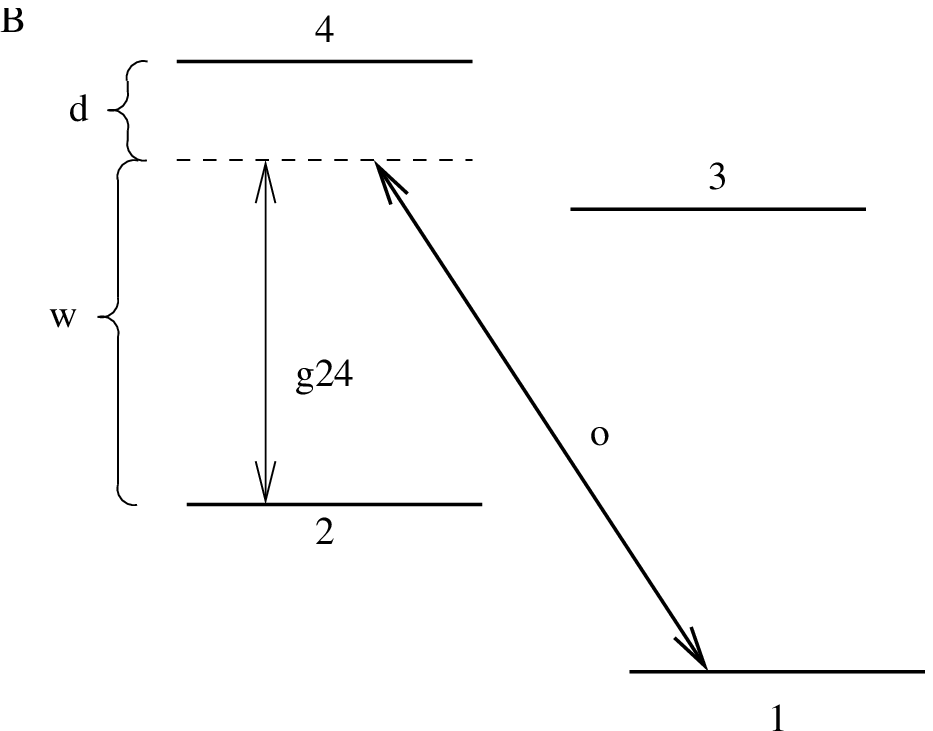}
\caption{\label{flippass} {\bf a}: Configuration of the $\pi / 2$-pulse. Two driving lasers in two-photon transition with identical Rabi frequencies $\Lambda$ couple to the atomic transitions $1 \leftrightarrow 3$ and $2 \leftrightarrow 3$. {\bf b}: Configuration for the STIRAP process. A driving laser couples to the $1 \leftrightarrow 4$ transition with Rabi frequency $\Theta$. The cavity mode couples to transitions $2 \leftrightarrow 4$ and $1 \leftrightarrow 3$, where the coupling to $1 \leftrightarrow 3$ is ineffective and not shown.}
\end{figure}

For two components, one can do the measurements as follows. First the external driving laser $\Omega$ is switched off. Then the roles of atomic levels 1 and 2 are interchanged in each atom via a Raman transition by applying a $\pi / 2$-pulse. To this end the transitions $1 \leftrightarrow 3$ and $2 \leftrightarrow 3$ are driven with two lasers (both have the same Rabi frequency $\Lambda$) in two-photon resonance for a time
$T = \pi \delta_{\Lambda} / |\Lambda|^2$ ($\delta_{\Lambda}$ is the detuning from atomic level 3). The configuration is shown in figure \ref{flippass}a. This pulse results in the mapping
$|1_j\ket \leftrightarrow |2_j\ket$ for all atoms $j$.

Next another laser, $\Theta$, that drives the transition
$1 \leftrightarrow 4$ is switched on, see figure \ref{flippass}b. Together with the coupling $g_{24}$, this configuration can be described in terms of three polaritons, $q_0^{\dagger}$, $q_+^{\dagger}$ and $q_-^{\dagger}$, in an analogous way to
$p_0^{\dagger}$, $p_+^{\dagger}$ and $p_-^{\dagger}$, where now the roles of the atomic
levels 1 and 2 and the levels 3 and 4 are interchanged.
Hence, if we choose $\Theta = \Omega$ the $\pi / 2$-pulse maps the $b^{\dagger}$ onto the dark state polaritons of the new configuration, $q_0^{\dagger}$, whereas
if we choose $\Theta = - \Omega$ it maps the $c^{\dagger}$ onto $q_0^{\dagger}$. The driving laser is then adiabatically switched off, $\Theta \rightarrow 0$, and the corresponding STIRAP process maps the $q_0^{\dagger}$ completely onto atomic excitations of level 1. This process can now be fast since the detuning $\Delta$ is significantly smaller than $\delta$ and hence the energies of all polariton species $q_0^{\dagger}$, $q_+^{\dagger}$ and $q_-^{\dagger}$ well separated. Another $\pi / 2$-pulse finally maps the excitations of level 1 onto excitations of level 2,
which can be measured by state selective resonance fluorescence in the same way as discussed in \cite{I02,HBP06}.

The whole sequence of $\pi / 2$-pulse, STIRAP process and another $\pi / 2$-pulse can be done much faster than the timescale set by the dynamics of the Hamiltonian (\ref{bosehubbard}) \cite{HBP06} and $b^{\dagger}$ or $c^{\dagger}$ can be mapped onto atomic excitations in a time in which they are not able to move between sites. 
The procedure thus allows to measure the instantaneous local particle statistics of each species separately.

\paragraph{Summary} We have shown that a two-component Bose-Hubbard model of polaritins can be created
in coupled arrays of high-Q cavities. As a new feature, the model can display transitions between the two
particle species. An experimental realization is feasible with cavities that have cooperativity factors much greater than unity and interact with the atoms for sufficient time. The local particle number statistics of both species can be measured independently with high accuracy.

This work is part
of the QIP-IRC supported by EPSRC (GR/S82176/0), the Integrated
Project Qubit Applications (QAP) supported by the IST directorate
as Contract Number 015848' and was supported by the EPSRC grant EP/E058256,
the Alexander von Humboldt Foundation, the Conselho Nacional de Desenvolvimento
Cient\'ifico e Tecnol\'ogico (CNPq) and the Royal Society.

\end{document}